\documentclass[aps,preprint,groupedaddress]{revtex4}
\usepackage{amsmath}
\usepackage{graphicx}
\begin{document}

\title{The Trivial Higgs at LHC}

\author{P. Cea$^{1,2}$}
\email[]{Paolo.Cea@ba.infn.it}
\author{L. Cosmai$^{1}$}
\email[]{Leonardo.Cosmai@ba.infn.it}
\affiliation{$^1$INFN - Sezione di Bari, I-70126 Bari, Italy\\
$^2$Physics Department, Univ. of Bari, I-70126 Bari, Italy }

\begin{abstract}
We further elaborate on  our proposal for the Trivial Higgs that within the Standard Model  is the unique possibility to implement the  spontaneous symmetry breaking of the local gauge symmetry by elementary local scalar fields. The Trivial Higgs boson turns out to be rather heavy  with mass  $m_H \simeq  750$ GeV. 
We  discuss the experimental signatures of our Trivial Higgs and compare with the recent data from ATLAS and CMS collaborations 
based on a total integrated luminosity between 1 fb$^{-1}$ and  2.3 fb$^{-1}$. We suggest that the available experimental data
could be consistent with our scenario.
\end{abstract}

%
\maketitle
\section{Introduction}
\label{s-1}
In a previous paper~\cite{Cea:2009} we have enlightened  the scenario where the Higgs boson without self-interaction (Trivial Higgs) could coexist with spontaneous symmetry breaking. 
Due to the peculiar rescaling of  the Higgs condensate, the relation between $m_H$ and the physical $v_R$ is not the same as in perturbation theory.  
According to this picture   one expects that  the ratio $m_H/v_R$ would  be a cutoff-independent constant.  In fact, our numerical results~\cite{Cea:2009}
 showed that the extrapolation to the continuum limit leads to the quite simple result:
\begin{equation}
\label{1.1}
m_H \; \simeq \;  \pi  \;  v_R  \; ,
\end{equation}
pointing to  a rather massive Trivial Higgs boson $m_H \; \simeq \; 750$ GeV. \\
\indent
A cornerstone of the Standard Model is the mechanism of spontaneous symmetry breaking that is mediated by the Higgs boson.
In fact, the discovery of the Higgs boson is the highest priority of the Large Hadron Collider (LHC).
Recently, both the ATLAS and CMS collaborations~\cite{Nisati:2011,Sharma:2011} reported the experimental results for the search of the  Higgs boson at the Large Hadron Collider
running at $\sqrt{s} = 7$ TeV, based on a total integrated luminosity between 1 fb$^{-1}$ and  2.3 fb$^{-1}$. \\
\indent
The aim of the present paper is to further elaborate on the production mechanisms of our Trivial Higgs and to compare the theoretical expectations with selected available data
from LHC.
\section{The Trivial Higgs}
\label{s-2}
For Higgs mass in the range $700 - 800 \; {\text{GeV}}$ the main production mechanism at LHC is the gluon fusion $gg  \rightarrow H$. The theoretical estimate of the production cross section at LHC for centre of mass  energy $\sqrt{s} = 7 \, {\text{TeV}}$ is~\cite{Dittmaier:2011ti} :
\begin{equation}
\label{2.1}
\sigma^{SM}(gg  \rightarrow H) \; \simeq \; 0.06  -  0.14  \; {\text{pb}} \; \; , \; \;  700 \; {\text{GeV}} \; < m_H \; < \; 800   \;  {\text{GeV}} \; .
\end{equation}
The gluon coupling to the Higgs boson in the Standard Model is mediated by triangular loops of top and bottom quarks. Since the Yukawa coupling of the Higgs particle to heavy quarks grows with quark mass, thus balancing the decrease of the triangle amplitude, the effective gluon coupling approaches a non-zero value for large loop-quark masses. On the other hand, we already 
argued~\cite{Cea:2009} that the non trivial rescaling of the Higgs condensate means that, if the fermions acquires a finite mass through the Yukawa couplings, then  the coupling of the physical Higgs field to the fermions could be very different from the perturbative Standard Model Higgs boson.  However, the coupling of the Higgs field to
the gauge vector bosons is fixed by the gauge symmetries, therefore the coupling of the Trivial Higgs boson to the  gauge vector bosons is the same as for the Standard Model Higgs boson.  
For large Higgs masses the vector-boson fusion mechanism becomes competitive to the gluon fusion Higgs production~\cite{Dittmaier:2011ti}:
\begin{equation}
\label{2.2}
\sigma^{SM}(W^+ \, W^-   \rightarrow H) \; \simeq \; 0.02  -  0.03  \; {\text{pb}} \; \; , \; \;  700 \; {\text{GeV}} \; < m_H \; < \; 800   \;  {\text{GeV}} \; .
\end{equation}
The main difficulty in the experimental identification of a very heavy  Higgs ($m_H > 650 \; {\text{GeV}}$)  resides in the  large width which makes impossible to observe a mass peak.  In fact, the expected  mass spectrum of our trivial Higgs  should be proportional to the Lorentzian distribution:
\begin{equation}
\label{2.3}
L_ H (E) \;  \sim  \;  \; \frac{\Gamma}{(E \; - \; 750 \; {\text{GeV}})^2 \; + \;  \Gamma^2}  \;   \;  \; , 
\end{equation}
where we assume the central value of the Higgs mass according to Eq.~(\ref{1.1}) and $\Gamma$ is the Higgs total width.
Note that  Eq.~(\ref{2.3}) is the simplest distribution consistent with the Heisenberg uncertainty principle and the finite lifetime
$\tau \simeq 1/\Gamma$. \\ 
According to  the triviality and spontaneous symmetry breaking scenario the Higgs self-coupling vanishes so that the decay width is mainly given by the  decays  into pairs of massive gauge bosons:
\begin{equation}
\label{2.4}
\Gamma  \; \simeq \; \Gamma( H \rightarrow W^+ \, W^-)  \; + \; \Gamma( H \rightarrow Z^0 \, Z^0) \; .
\end{equation}
Since the coupling of the Trivial Higgs to the gauge vector bosons is fixed by the local gauge symmetry, the decay width into gauge vector bosons is the same as in the case of the perturbative Higgs boson, consequently~\cite{Djouadi:2005gi}
\begin{equation}
\label{2.5}
 \Gamma( H \rightarrow W^+ \, W^-)   \; \simeq \;  2 \; \Gamma( H \rightarrow Z^0 \, Z^0) \;  \; \sim  \; \; G_F \; m_H^3 \; .
\end{equation}
Equation~~(\ref{2.5}) shows that in the high mass region $m_H \gtrsim 400$ GeV the Higgs total width depends strongly on
$m_H$. To take care of the energy dependence of the width in Eq.~(\ref{2.3}) we need $\Gamma(E)$ as a function of E. To this end
we may follow the calculations of Higgs total width within the Standard Model performed in Ref.~\cite{Dittmaier:2011ti}.
\begin{figure}[t]
\includegraphics[width=0.47\textwidth,clip]{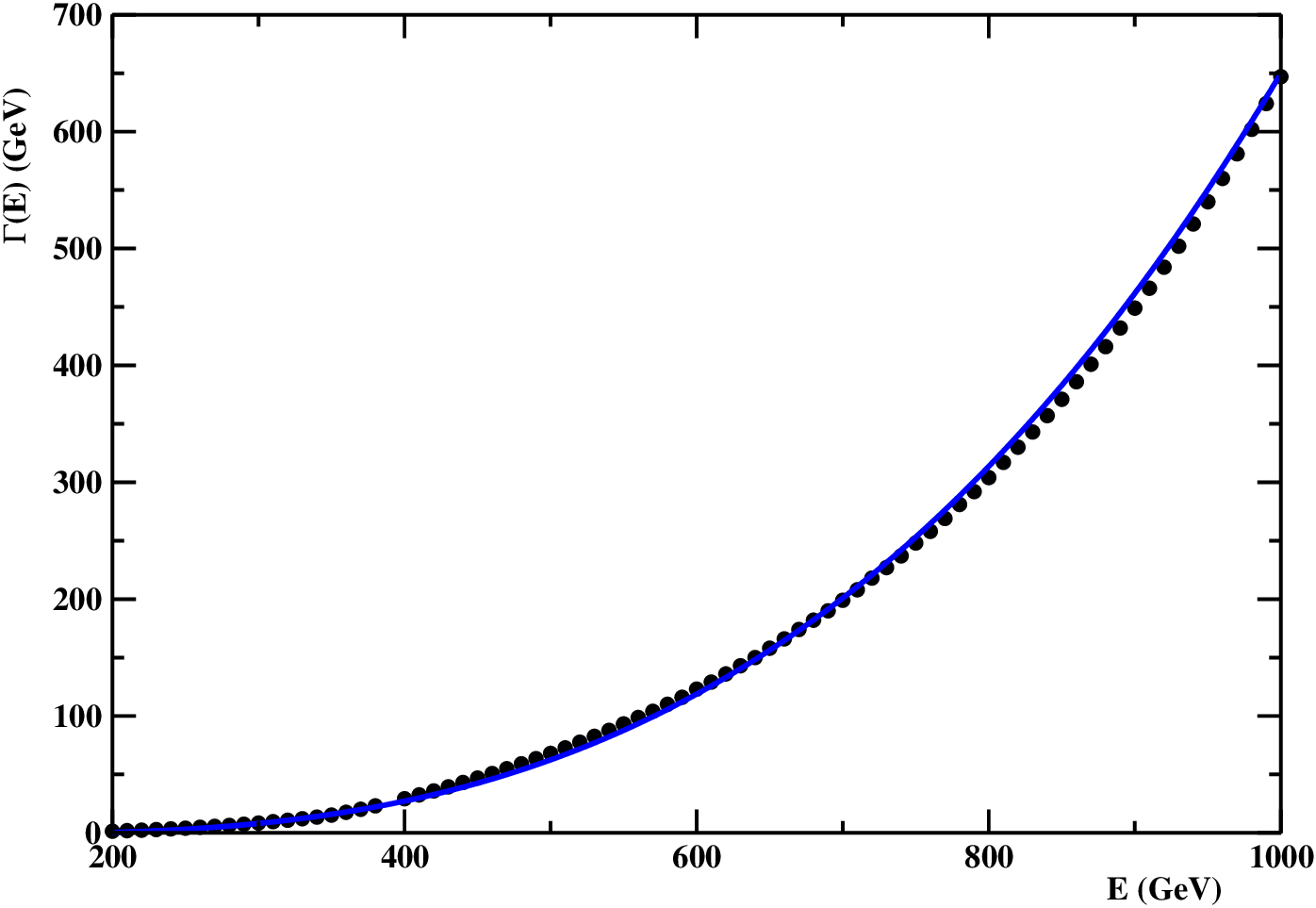}
\includegraphics[width=0.47\textwidth,clip]{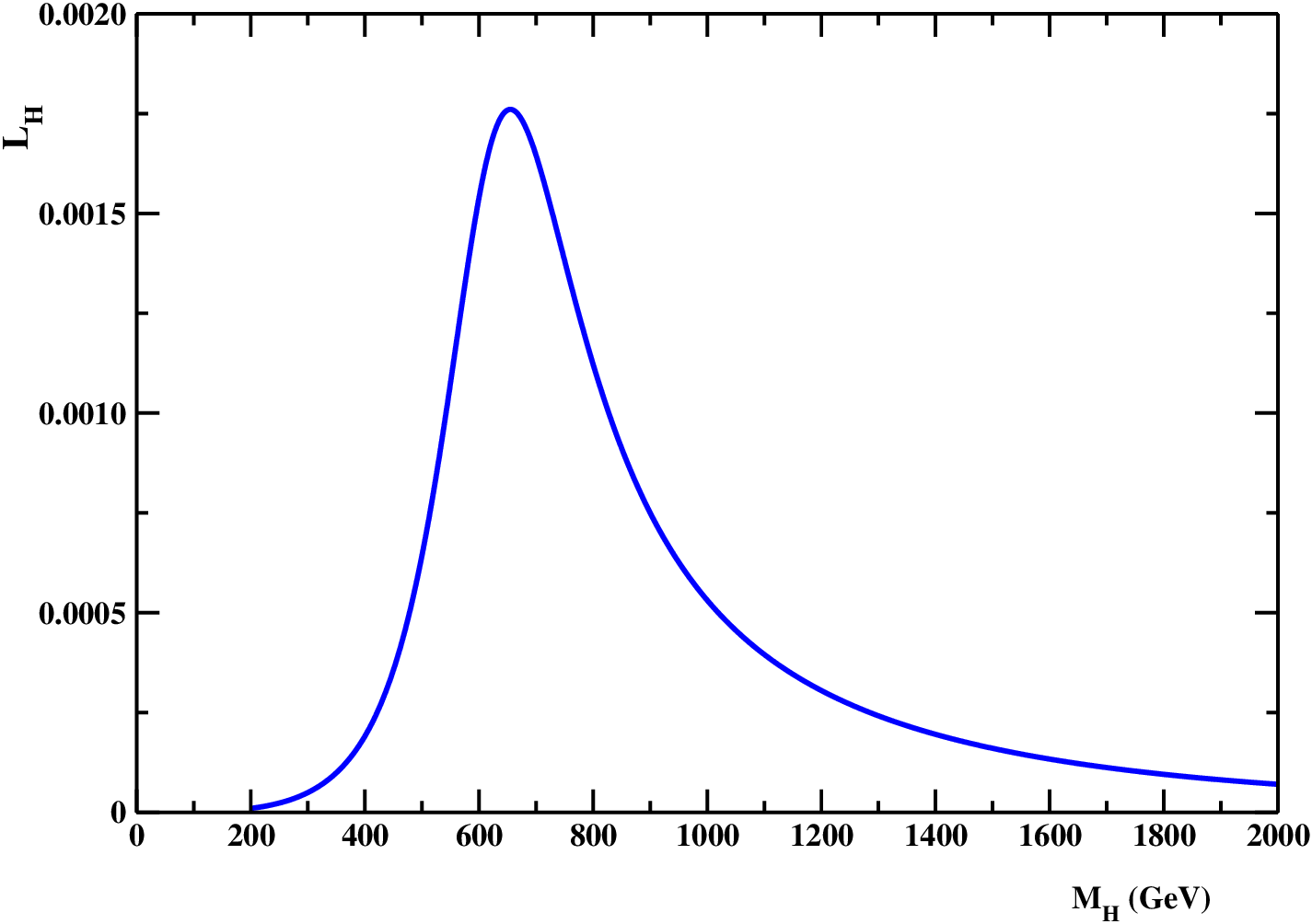}
\caption{\label{fig-1-2} (Left) The energy dependence of the Higgs total width.  Full points are the values of the Standard Model Higgs total widths reported in Tables 28 and 29 of Ref.~\cite{Dittmaier:2011ti}. The full line is our parameterization  Eq.~(\ref{2.6}).
(Right) The Lorentzian distribution Eq.~(\ref{2.7}) as a function of the Higgs mass $M_H$ .}
\end{figure}
In  Fig.~\ref{fig-1-2} (left) we show the energy dependence of the Higgs total width $\Gamma( E)$ according to Tables 28 and 29 of Ref.~\cite{Dittmaier:2011ti}. We have fitted the tabulated values of $\Gamma( E)$ with the phenomenological relation (full blue line 
in Fig.~\ref{fig-1-2}, left):
\begin{equation}
\label{2.6}
\Gamma( E)  \; \simeq \; 8.0 \; 10^{-7} \; E^3 \; - \; 1.5 \; 10^{-4} \; E^2 \; .
\end{equation}
We obtain, therefore, the following Lorentzin  distribution:
\begin{equation}
\label{2.7}
L_ H (E) \;  \simeq  \;  \; \frac{1.465}{\pi} \; \frac{\Gamma(E)}{(E \; - \; 750 \; {\text{GeV}})^2 \; + \;  \Gamma(E)^2} \;  \;   \;  \; , 
\end{equation}
where $\Gamma( E)$ is given by Eq.~(\ref{2.6}), and the normalization is  such that:
\begin{equation}
\label{2.8}
\int^{\infty}_{0} \; L_ H (E)  \;  dE \; \; = \; \; 1 \; \; .
\end{equation}
In the limit $\Gamma \; \rightarrow \; 0$,  $L_ H (E)$ reduces to $\delta(E \; - 750 \; {\text{GeV}})$.  In Fig.~\ref{fig-1-2} (right) we display  the Lorentzian distribution  as a function of the energy E. \\
To evaluate the Higgs event production at LHC we need the Higgs production total cross section. As already discussed, 
for large Higgs masses the main production processes are by vector-boson fusion and gluon-gluon fusion. The Trivial Higgs production cross section by vector-boson fusion is almost the same as in the perturbative Standard Model calculations.
\\
In Fig.~\ref{fig-3-4} (left) we show the energy dependence of the perturbative Higgs boson vector-boson fusion cross section 
 within the Standard Model  at $\sqrt{s}$ = 7 TeV reported in Table 11 of Ref.~\cite{Dittmaier:2011ti}. We parametrize the
 energy dependence of the cross section as (full blue line in Fig.~\ref{fig-3-4}, left):
\begin{equation}
\label{2.9}
\sigma^{SM}(W^+ \, W^-   \rightarrow H) \; \simeq \;    \left (  \frac{ 3.0 \; 10^5}{ M_H} 
 \; - \; \frac{8.0 \; 10^6}{ M_H^2}  \right )  \;  \exp (- 0.0035 M_H) 
\; \; , \; \;  M_H \; {\text{ in \;  GeV}}  \; .
\end{equation}
As concern the gluon fusion cross section, in Fig.~\ref{fig-3-4} (right) we display the  Standard Model 
Higgs production cross section at $\sqrt{s}$ = 7 TeV reported in Table 5 of Ref.~\cite{Dittmaier:2011ti}. Again we parametrize the
 energy dependence of the cross section as (full blue line in Fig.~\ref{fig-3-4}, right):
\begin{equation}
\label{2.10}
\sigma^{SM}(gg  \rightarrow H) \; \simeq \;  
 \left\{ \begin{array}{ll}
 \;  \left (  \frac{ 1.1 \; 10^7}{ M_H} 
 \; + \;0.0097 \; M_H^3  \right )  \;  \exp (- 0.016 M_H)  \; \; &  M_H \; \leq \; 320 \; 
  \\
 \; \; \; \;   2.25 \; 10^3  \; \; \; \; & 320 \; \leq \; M_H \; \leq \; 380
  \\
 \; \;  \; \; 2.30 \; 10^3 \;  \;  \exp (- 0.016 M_H)  \; \; &  380 \; \leq \; M_H
\end{array}
    \right.
\end{equation}
where $M_H$  is expressed in  GeV. \\
\begin{figure}[t]
\includegraphics[width=0.47\textwidth,clip]{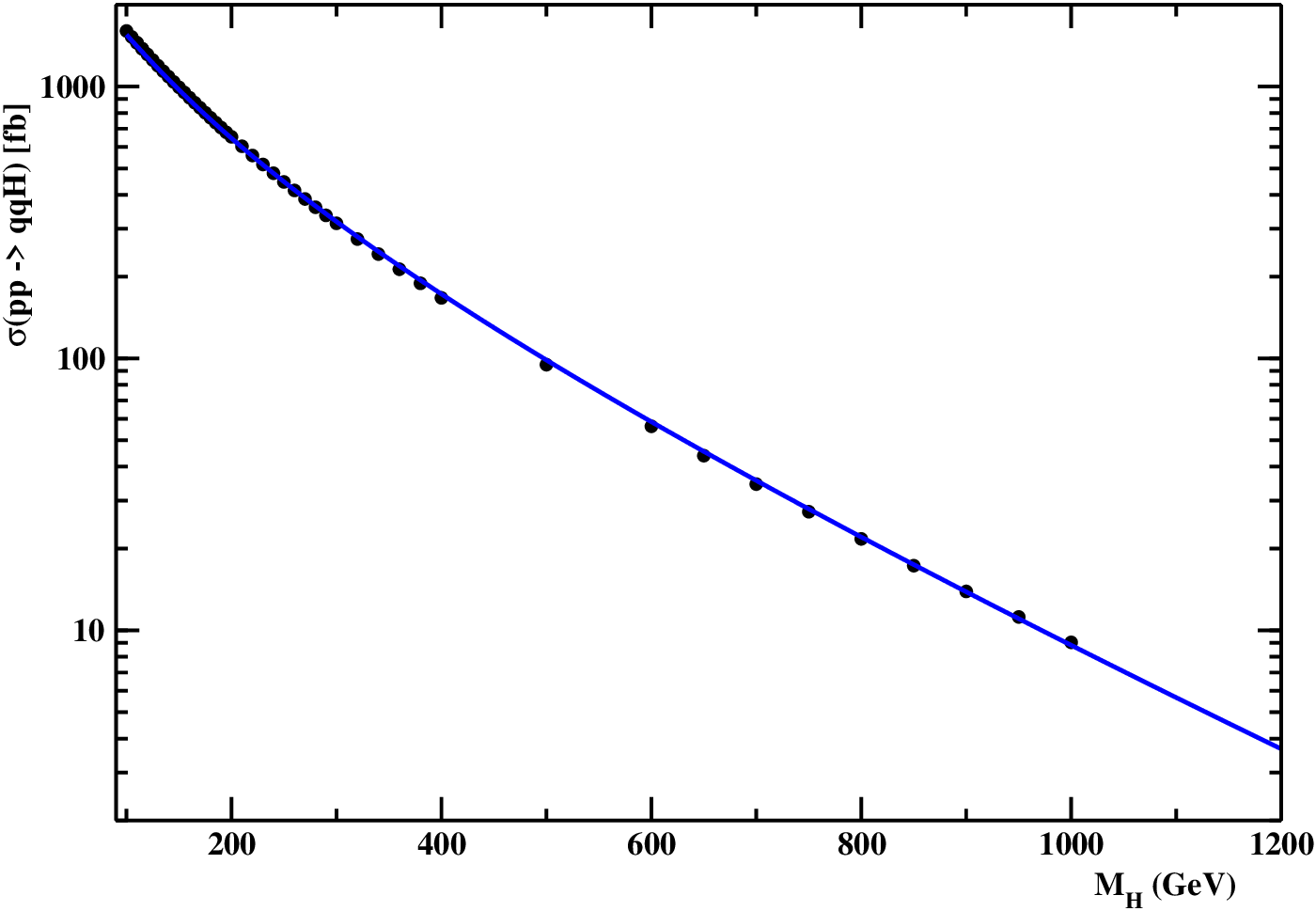}
\includegraphics[width=0.47\textwidth,clip]{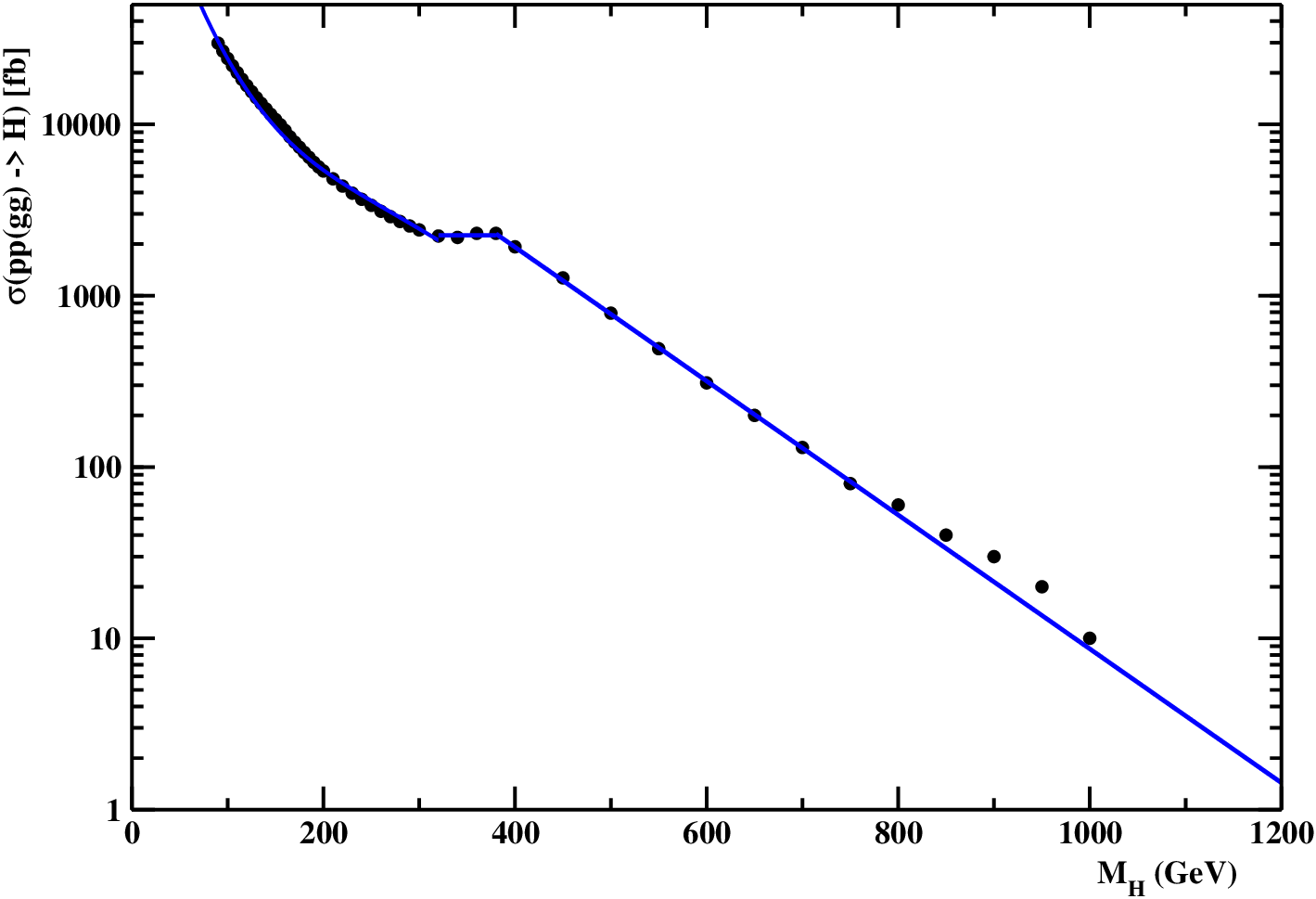}
\caption{\label{fig-3-4} (Left) The energy dependence of the  vector-boson fusion cross section. The full points are the Standard Model  Higgs production cross section at $\sqrt{s}$ = 7 TeV reported in Table 11 of Ref.~\cite{Dittmaier:2011ti}. The full line is our parameterization  Eq.~(\ref{2.9}).
(Right) The energy dependence of the  gluon-fusion cross section. The full points are the Standard Model 
Higgs production cross section at $\sqrt{s}$ = 7 TeV reported in Table 5 of Ref.~\cite{Dittmaier:2011ti}. The full line is our parameterization  Eq.~(\ref{2.10}).}
\end{figure}
The gluon coupling to the perturbative Higgs boson in the Standard Model is mediated by triangular loops of top and bottom quarks. On the other hand,  as already discussed, the coupling of the Trivial Higgs to the fermions  is not fixed by the local gauge symmetry so that there is no a priori reasons to expect that these couplings would be proportional to the fermion masses. However, in the relevant high mass region $m_H \gtrsim 400$ GeV the energy dependence of $\sigma^{SM}(gg  \rightarrow H)$ is given essentially by the gluon distribution  function. Thus, we may safely assume that the Trivial Higgs production cross section by gluon-gluon fusion is proportional to $\sigma^{SM}(gg  \rightarrow H)$  as given by Eq.~(\ref{2.10}). \\
To summarize, our main approximation for the total production cross section of the Trivial Higgs is:
\begin{equation}
\label{2.11}
\sigma(p \; p \;  \rightarrow \; H \; + \; X) \; \simeq \;    \sigma^{SM}(W^+ \, W^-   \rightarrow H)
\; + \; \kappa \; \sigma^{SM}(gg  \rightarrow H) \; ,
\end{equation}
where the parameter $\kappa$ takes care of our ignorance on the Yukawa couplings of the Trivial Higgs to the fermions.\\
To compare the invariant mass spectrum of our Trivial Higgs with the experimental data, we observe that:
\begin{equation}
\label{2.12}
 N_H (E_{1},E_{2} )  \; \simeq \; {\cal{L}} \;  \int^{E_2}_{E_1} \; BR(E) \;  \varepsilon(E) \; \sigma(p \; p \;  \rightarrow \; H \; + \; X)  \; L_ H (E)  \;  dE  
 \;  \; , 
\end{equation}
where  $N_ H$ is the number of Higgs events in the energy interval $E_1,E_2$, corresponding to an integrated luminosity 
${\cal{L}}$, in the given channel with branching ratio $BR(E)$.  
The parameter $ \varepsilon(E)$  accounts for  the efficiency of trigger, acceptance of the detectors, the kinematic selections, and so on. Thus, in general $ \varepsilon(E)$ depends on the energy, the selected channel and the detector. In our preliminary study we
shall adopt the rather crude approximation: 
\begin{equation}
\label{2.13}
 \varepsilon(E) \; \simeq \;  0.20   \;  \; .
\end{equation}
As concern the branching ratios BR(E), in the relevant Higgs mass region $m_H \gtrsim 400$ GeV the Higgs decays mainly into
pairs of massive gauge bosons.  The most important decay channels are $H  \rightarrow WW  \rightarrow \ell \nu q q$,  $H  \rightarrow ZZ   \rightarrow \ell \ell q q$,  $H  \rightarrow ZZ  \rightarrow \ell \ell \nu \nu$ and 
$H  \rightarrow ZZ  \rightarrow \ell \ell  \ell \ell$.
Since the relevant branching ratios are almost independent on the Higgs mass and, assuming the Standard Model values, we are led to use the following values:
\begin{equation}
\label{2.14}
  \begin{array}{ll}
BR(H  \rightarrow WW  \rightarrow \ell \nu q q ) \; \; \simeq  \;  \; \; \; \;  0.438 \; \; 
  \\
BR(H  \rightarrow ZZ   \rightarrow \ell \ell q q ) \;  \; \; \; \;  \simeq  \;  \; \; \; \;  0.153 \; \; 
  \\
BR(H  \rightarrow ZZ  \rightarrow \ell \ell \nu \nu ) \; \; \;  \;  \simeq  \;  \; \; \; \; \, 0.061 \; \; 
 \\
BR(H  \rightarrow ZZ  \rightarrow \ell \ell  \ell \ell ) \;  \; \; \; \;  \simeq  \;  \; \; \; \; \, 0.010 \; \; 
\end{array}
\end{equation}
The above approximations should be sufficient for the preliminary study of interest here.
\section{$H  \rightarrow WW   \rightarrow \ell \nu q q $}
\label{s-3}
\begin{figure}
\includegraphics[width=0.47\textwidth,clip]{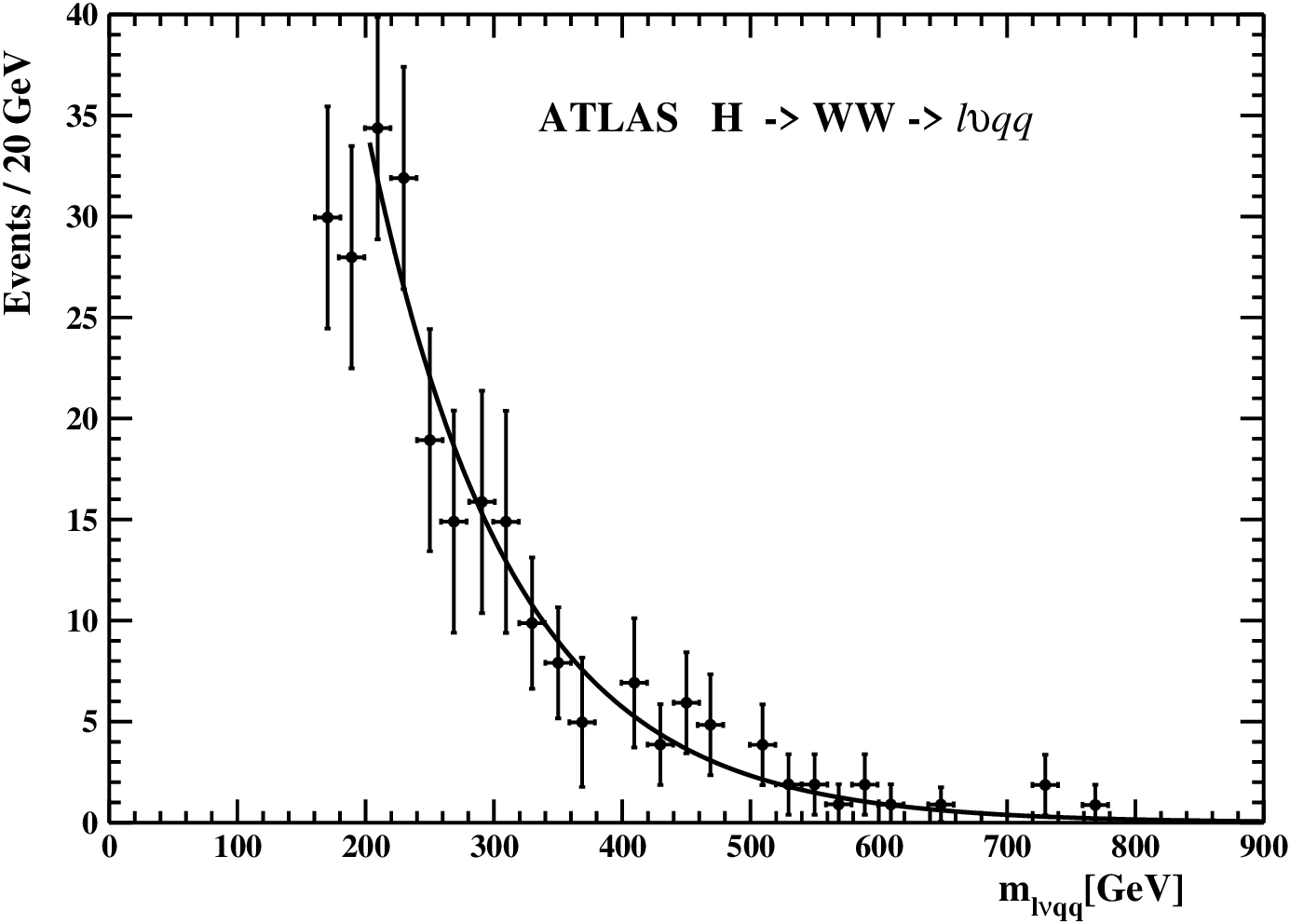}
\includegraphics[width=0.47\textwidth,clip]{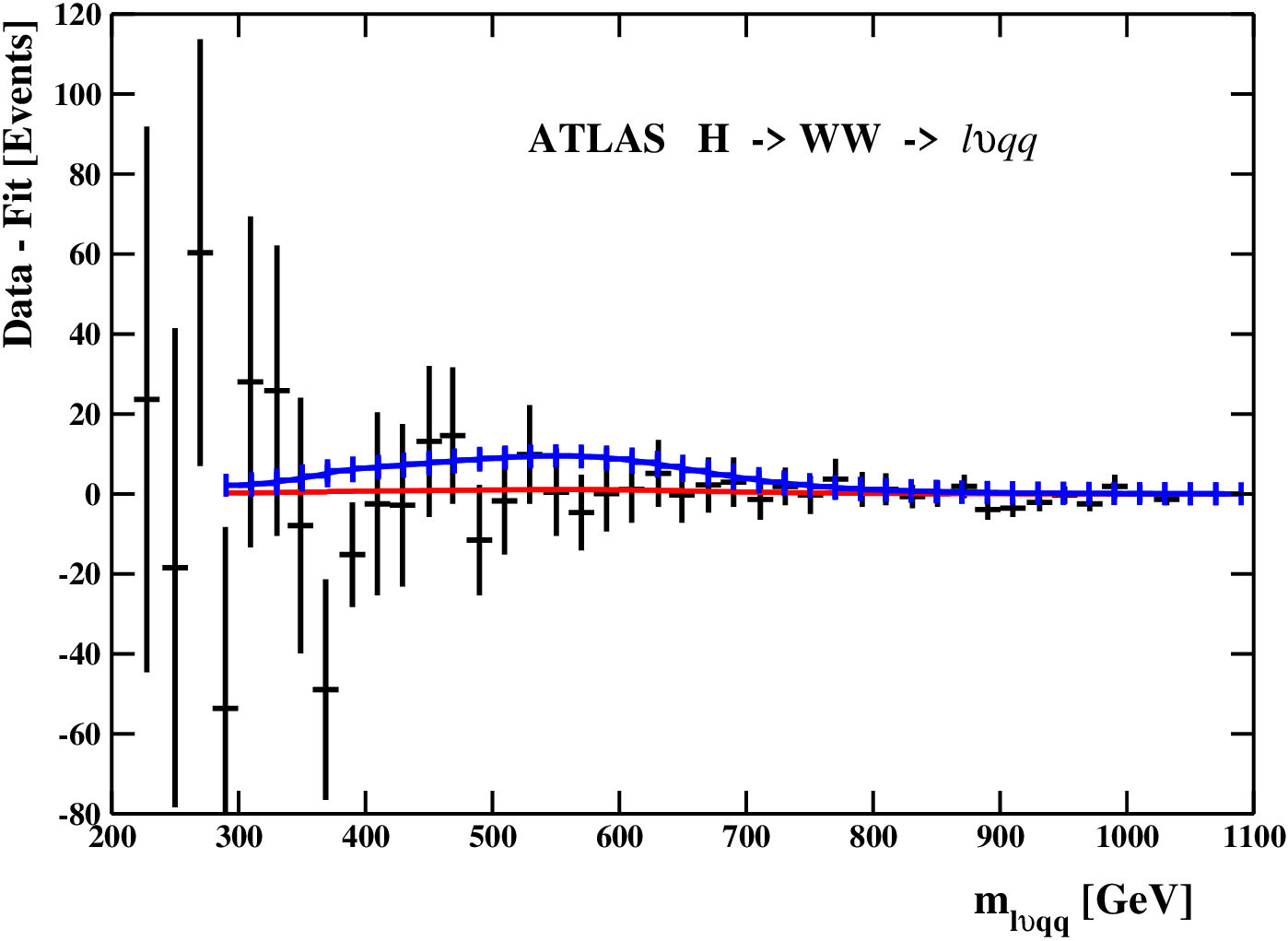}
\caption{\label{fig-5-6} (Left) Distribution of the invariant mass $m_{\ell \nu qq}$ for the  process $H \; \rightarrow WW \; \rightarrow \ell \nu qq$ corresponding to an integrated luminosity of 35 pb$^{-1}$. The data has been extracted from Fig. 4, panel b) of 
Ref.~\cite{ATLAS:2011a}. The  invariant mass continuum  background is parametrized as a  falling exponential function (full line). \\
(Right) Comparison of the background subtracted experimental data  corresponding to an integrated luminosity of 1.04 
fb$^{-1}$ (data taken from Ref.~\cite{Murray:2011}) with the Higgs event distribution according to Eq.~(\ref{2.12})  binned in energy intervals of 20 GeV assuming $\kappa = 1$ (red line) and $\kappa = 10$ (blue line).}
\end{figure}
For high Higgs mass the decay process $H  \rightarrow WW  \rightarrow \ell \nu q q$ has the largest branching ratio. Moreover,
the presence of charged lepton allows to obtain a good rejection of the QCD processes. The main background is given by the production of W + jet which, however, should be suppressed for large invariant mass $m_{\ell \nu qq}$. \\
Preliminary results from  the ATLAS collaboration~\cite{ATLAS:2011a}  reported the experimental  search of the  Higgs boson at the Large Hadron Collider running at $\sqrt{s} = 7$ TeV, based on a total integrated luminosity of about 40 pb$^{-1}$. In particular, in Fig.~\ref{fig-5-6} (left) we display the distribution of the invariant mass for the
Higgs boson candidates corresponding to the process $H \; \rightarrow WW \; \rightarrow \ell \nu qq$. According to 
Ref.~\cite{ATLAS:2011a}, the events were selected requiring exactly one lepton with $p_T \; > 30$ GeV. The missing transverse energy in the event were required to be $E^{miss}_T \; > \; 30$ GeV. The  invariant mass continuum  background is parametrized as a  falling exponential function. It is amusing to see that there are some events in the high invariant mass region which, however, are
not statistically significant due to the low integrated luminosity.
Recently, an update of search for the Higgs boson in this channel from the ATLAS collaboration corresponding to  an integrated luminosity of 1.04  fb$^{-1}$  have been presented in Ref.~\cite{Murray:2011}.  In Fig.~\ref{fig-5-6} (right)  we display the background-subtracted data. The large statistical uncertainties, mainly due to the background subtraction, do not show evidence of any structure in the invariant mass distribution. \\
To compare our theoretical expectation we use Eq.~(\ref{2.12})  with the branching ratios in Eq.~(\ref{2.14}). 
In Fig.~\ref{fig-5-6} (right) we compare the the Higgs event distribution  binned in energy intervals of 20 GeV assuming 
$\kappa = 1$ (red line), i.e. the Standard Model perturbative Higgs production cross section, and integrated luminosity of 
1.04  fb$^{-1}$. We see that our theoretical distribution is compatible with experimental data.
Interestingly enough, we find that the experimental data allow an enhanced gluon-gluon fusion cross section, as can be inspected 
in Fig.~\ref{fig-5-6} (right) where we also display the  theoretical Higgs event distribution  assuming  the total production cross section of the Trivial Higgs is given by  Eq.~(\ref{2.11}) with $\kappa = 10$ (blue line). \\
In the following Sections we consider the most important Higgs decay channels   in order
to determine if the enhanced Higgs production by the gluon-gluon fusion process is compatible with
available experimental observations.
\section{$H \rightarrow ZZ   \rightarrow \ell \ell q q $}
\label{s-4}
\begin{figure}
\includegraphics[width=0.47\textwidth,clip]{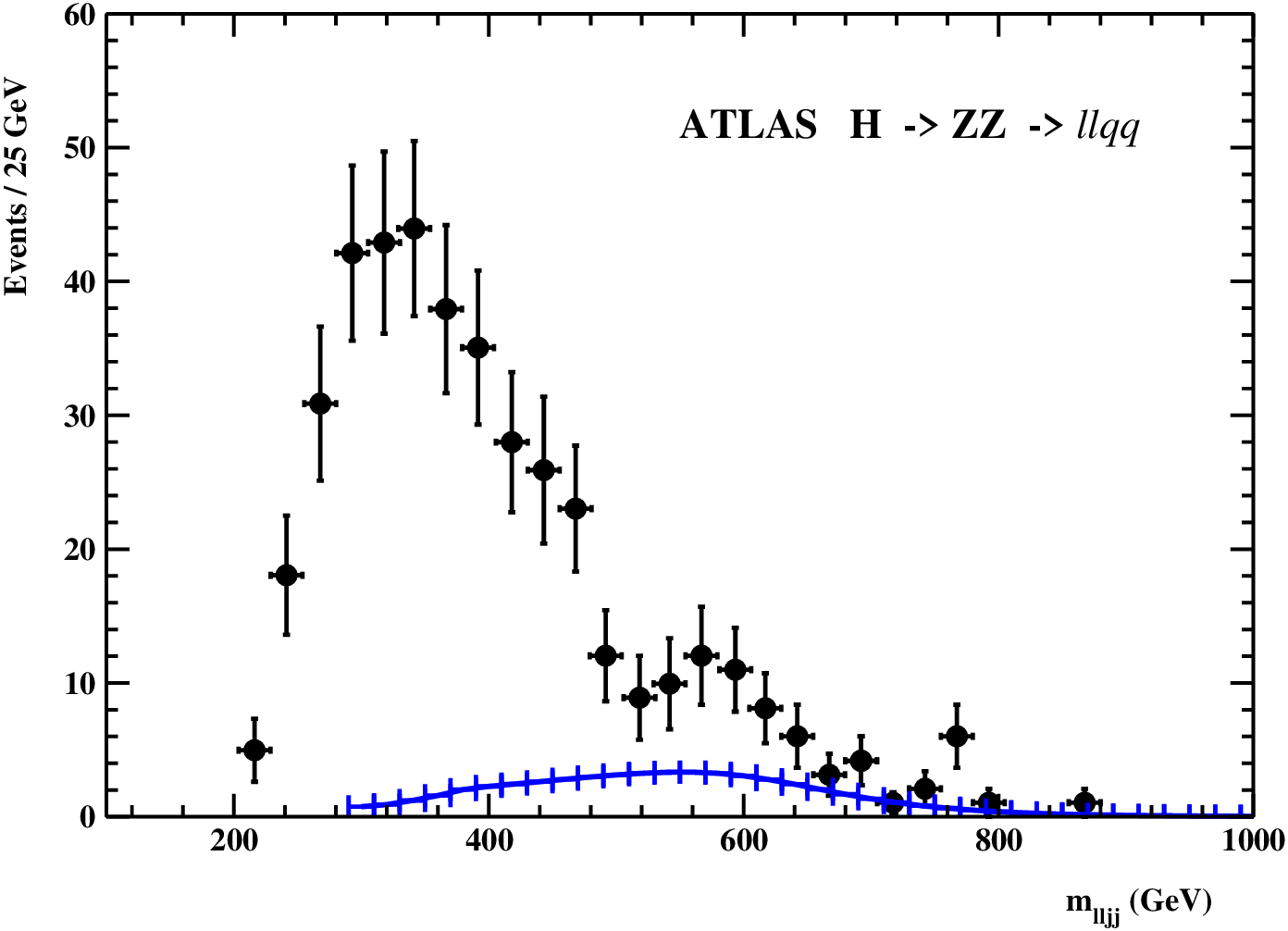}
\includegraphics[width=0.47\textwidth,clip]{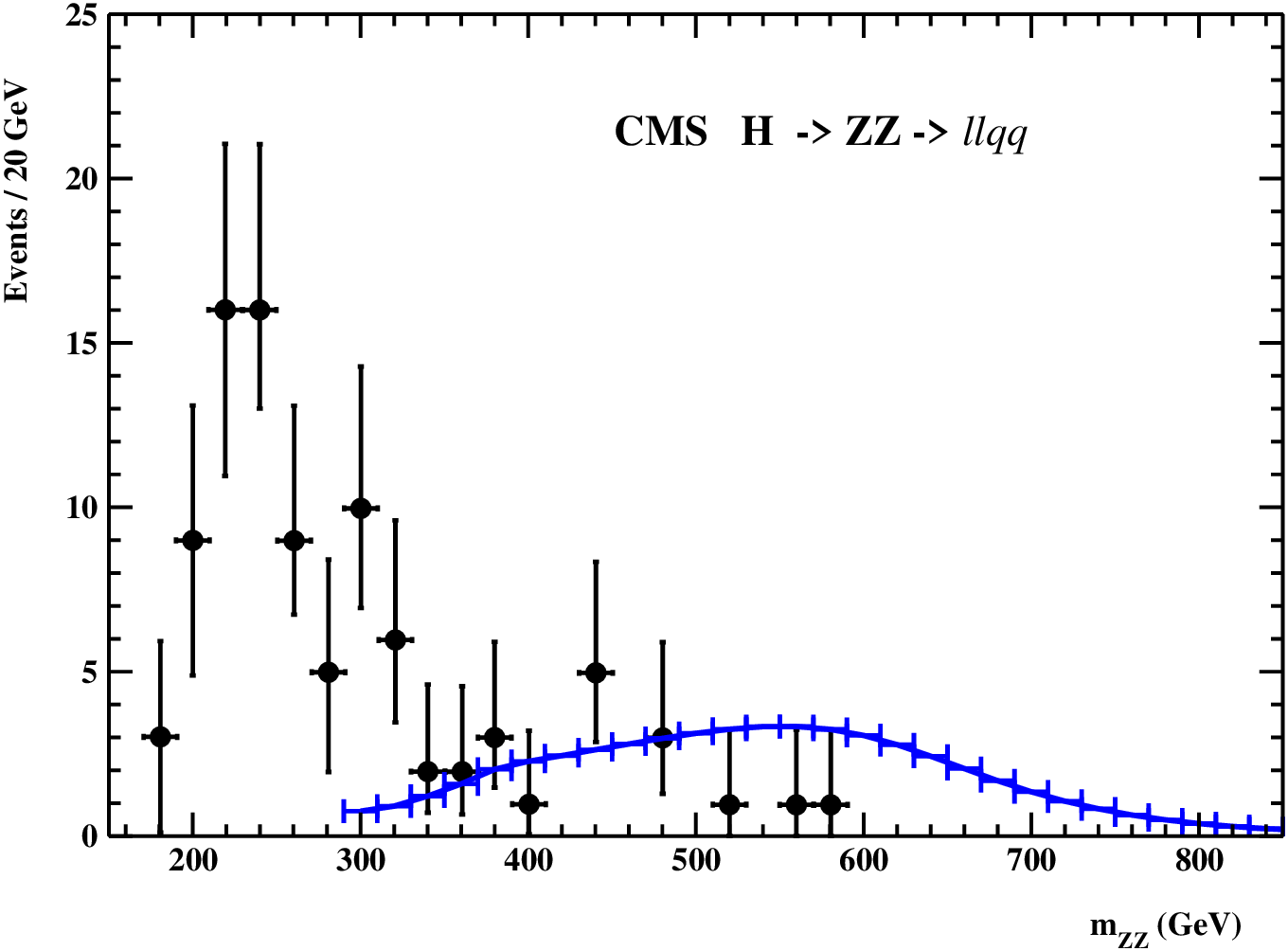}
\caption{\label{fig-7-8} (Left) Comparison of the distribution of the invariant mass $m_{\ell \ell qq}$ for the  process $H \; \rightarrow ZZ \; \rightarrow \ell \ell qq$ corresponding to an integrated luminosity of 1.04 fb$^{-1}$ with the Higgs event distribution according to Eq.~(\ref{2.12})  binned in energy intervals of 20 GeV assuming  $\kappa = 10$ (blue line). The data from the ATLAS Collaboration have been extracted from Fig.~2, panel c) of Ref.~\cite{ATLAS:2011b}. \\
(Right) Comparison of the distribution of the invariant mass $m_{ZZ}$ for the  process $H \; \rightarrow ZZ \; \rightarrow \ell \ell qq$ corresponding to an integrated luminosity of 1.6 fb$^{-1}$ with the Higgs event distribution according to Eq.~(\ref{2.12})  binned in energy intervals of 20 GeV assuming  $\kappa = 10$ (blue line). The data from the CMS Collaboration have been extracted from  Ref~\cite{Sharma:2011}. }
\end{figure}
The decay channel  $H \rightarrow ZZ   \rightarrow \ell \ell q q$ has the highest rate among all the processes where the Higgs boson
decays into two Z bosons. The search strategy is to find some structures in the invariant mass  $m_{\ell \ell qq}$. The major background is due to processes with production of Z + jet. In Fig.~\ref{fig-7-8} we report the experimental data from ATLAS (left) and
CMS (right) collaborations. Our theoretical expectations obtained from  Eq.~(\ref{2.12})  are compared with the data
(compare also with Fig.~10, panel c) of Ref.~\cite{CMS:2011a}). We see that our Higgs event distribution is not in contradiction
with the experimental data, even though we cannot exclude that the data are compatible with the background-only hypothesis.
\section{$H  \rightarrow ZZ  \rightarrow \ell \ell \nu \nu $ \& $H  \rightarrow ZZ  \rightarrow \ell \ell  \ell \ell  $}
\label{s-5}
\begin{figure}
\includegraphics[width=0.47\textwidth,clip]{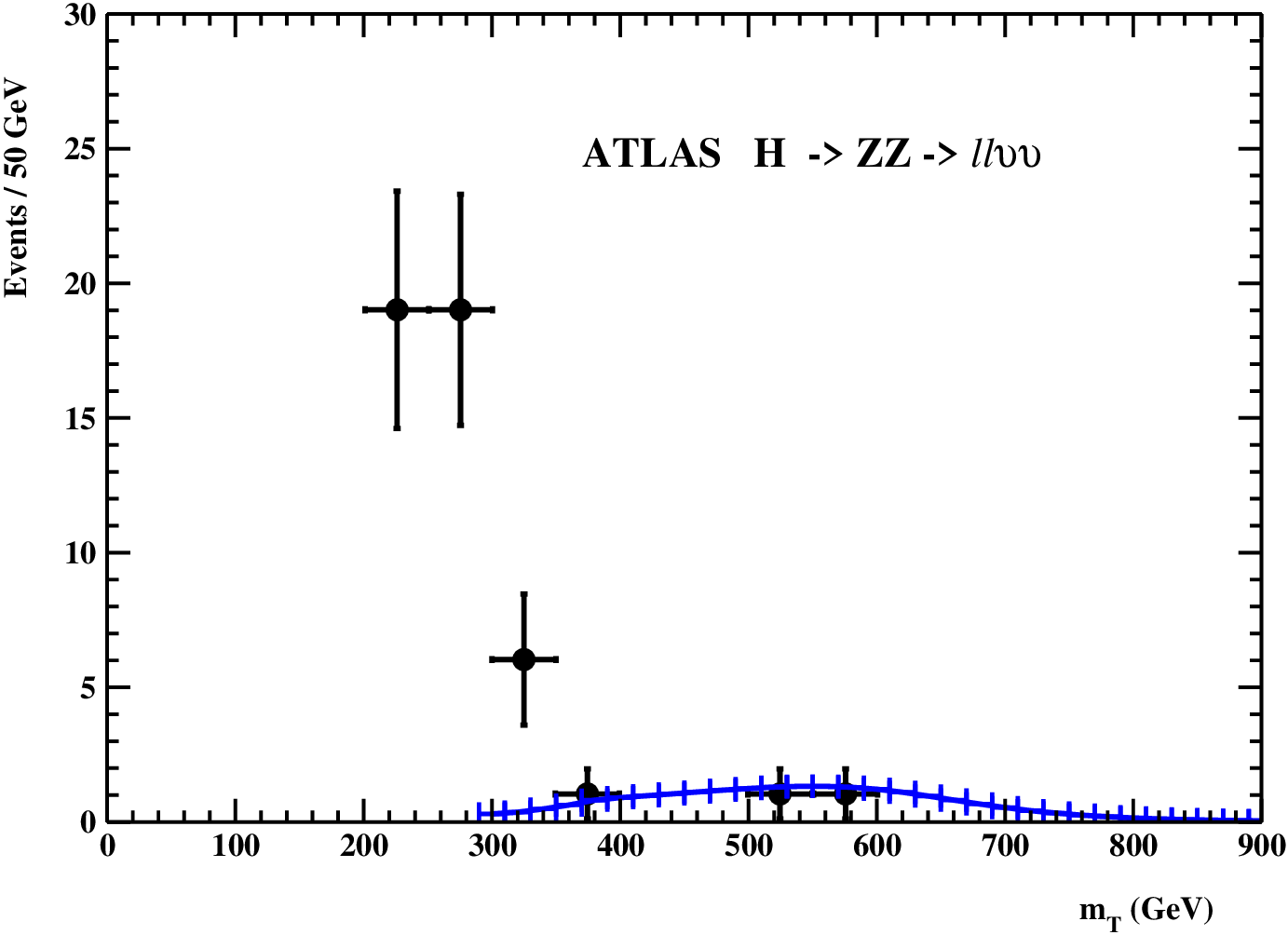}
\includegraphics[width=0.47\textwidth,clip]{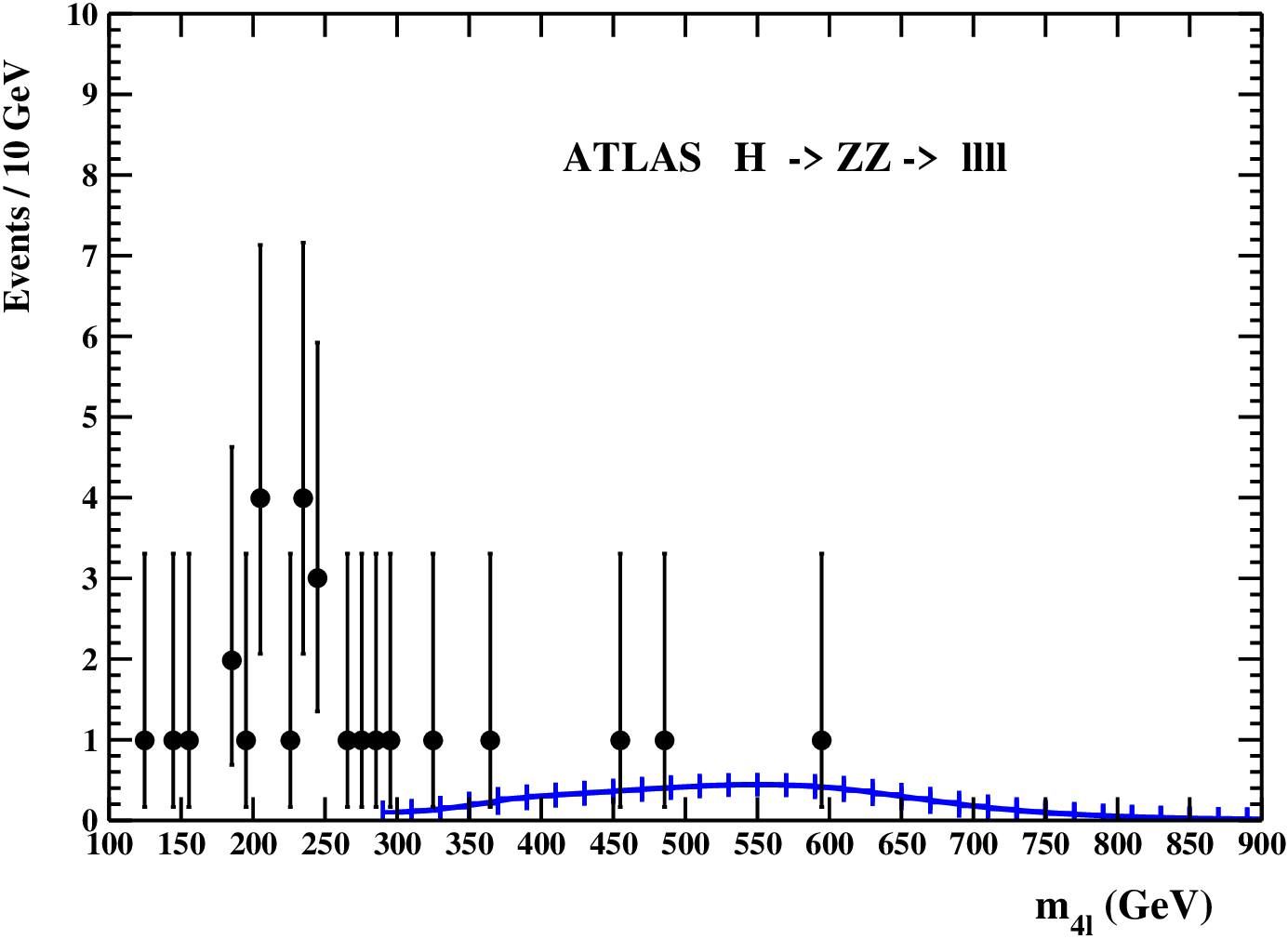}
\caption{\label{fig-9-10} (Left) Comparison of the distribution of the transverse mass $m_{T}$ for the  $H \; \rightarrow ZZ \; \rightarrow \ell \ell \nu \nu$ channel corresponding to an integrated luminosity of 1.04 fb$^{-1}$ with the Higgs event distribution  Eq.~(\ref{2.12})  binned in energy intervals of 20 GeV assuming  $\kappa = 10$ (blue line). The data from the ATLAS Collaboration have been extracted from Fig.~2, panel e) of Ref.~\cite{ATLAS:2011b}. \\
(Right) Comparison of the distribution of the invariant mass $m_{4 \ell}$ for the  process $H \; \rightarrow ZZ \; \rightarrow \ell \ell \ell \ell$ corresponding to an integrated luminosity of 1.96 - 2.28  fb$^{-1}$  with the Higgs event distribution according to Eq.~(\ref{2.12})  binned in energy intervals of 20 GeV assuming  
$\kappa = 10$ and ${\cal{L}} \simeq 2.0$ fb$^{-1}$  (blue line). The data  have been extracted from  Fig.~1, panel b) of 
Ref.~\cite{ATLAS:2011b}.}
\end{figure}
The channels $H  \rightarrow ZZ  \rightarrow \ell \ell \nu \nu $  and  $H  \rightarrow ZZ  \rightarrow \ell \ell  \ell \ell $ have the
lowest branching ratios, see Eq.~(\ref{2.14}). Nevertheless, the presence of leptons allows to efficiently reduce the huge 
background due mainly to diboson production. \\
In Fig.~\ref{fig-9-10} (left) we display the distribution of the  missing transverse energy for the channel 
 $H \; \rightarrow ZZ \; \rightarrow \ell \ell \nu \nu$ corresponding to an integrated luminosity of 1.04 fb$^{-1}$. The data
 from the ATLAS collaboration have been taken from  Ref.~\cite{ATLAS:2011b}. In the high mass region 
 $m_T \gtrsim 400$ GeV, where the background is strongly suppressed, there are a few events which compare well
 with our theoretical prediction assuming an enhanced gluon fusion cross section, $\kappa$ = 10. \\
 In Fig.~\ref{fig-9-10} (right) we  report the invariant mass distribution for the golden channel 
 $H \; \rightarrow ZZ \; \rightarrow \ell \ell \ell \ell$ with integrated luminosity 1.96 - 2.28 fb$^{-1}$ from the ATLAS
 collaboration~\cite{ATLAS:2011b}.  Even in this channel there are a few events in the high invariant mass region
 which are in fair agreement with our Higgs event distribution with   $\kappa$ = 10 and ${\cal{L}} \simeq 2.0$ fb$^{-1}$.
\section{Conclusion}
\label{s-7}
To conclude,  we argued that strictly  local scalar fields are compatible with spontaneous symmetry breaking. 
We stress that within the Standard Model our proposal is the unique possibility to implement the 
 spontaneous symmetry breaking of the local gauge symmetry by elementary local scalar fields. 
Our Trivial Higgs boson turns out to be rather heavy.  We compared our proposal with the recent results from ATLAS and
CMS collaborations and  gave some evidence that experimental data are not in contradiction with our scenario. Of course, it should be emphasized
that the data could be compatible with the background-only hypothesis.  We suggested that the Yukawa couplings of the
Trivial Higgs could be very different from the perturbative Standard Model Higgs boson. In particular, there are
no compelling reasons to expect that the couplings of the Trivial Higgs to fermions are proportional to the
fermion mass. As a consequence, we do not have enough arguments to set constraints on the Higgs production
by the gluon-gluon fusion processes. Nevertheless, it is remarkable that the available
data allow us to set an upper limit $\kappa  \lesssim 10$ to the gluon fusion Higgs production cross section. \\
\indent
We are confident that in the near future forthcoming data from LHC will confirm our Trivial Higgs proposal.
In the most favourable case of an enhanced gluon fusion cross section $\kappa \lesssim 10$ an integrated
luminosity of order 10 fb$^{-1}$ will be enough to confirm our scenario. On the other hand, if the gluon
fusion cross section turns out to be suppressed with respect to the Standard Model perturbative Higgs boson,
 $\kappa \lesssim 1$, then we must wait for an integrated luminosity of several hundred of fb$^{-1}$.

\end{document}